\documentclass[preprint,12pt]{elsarticle}

\usepackage{amssymb}
\usepackage{amsmath}

\usepackage{booktabs}
\usepackage{booktabs}
\usepackage{times}
\usepackage{url}
\usepackage{float}
\usepackage{algorithm,algorithmic}

\usepackage{bm}
\usepackage{mathrsfs}

\usepackage{amsmath}
\usepackage{longtable}
\usepackage{amsfonts}
\usepackage{amssymb}

\usepackage{graphicx}
\usepackage{epstopdf}

\usepackage{multirow}
\usepackage{subcaption}
\usepackage{verbatim}
\usepackage{lineno}
\usepackage{enumitem}
\usepackage{color}
\usepackage{subcaption}
\usepackage{epsfig}

\usepackage{lineno}
\modulolinenumbers[5]

\usepackage{bbm} 
\usepackage[T1]{fontenc}
\usepackage{array} 

\usepackage{ragged2e}

\usepackage{hyperref}
\usepackage{xcolor}

\journal{}

\begin{document}

\begin{frontmatter}

\title{Improved visual-information-driven model for crowd simulation and its modular application}

\author[mymainaddress]{Xuanwen Liang}
\author[2]{Jiayu Chen}
\author[mymainaddress]{Eric Wai Ming Lee}
\author[3]{Wei Xie\corref{mycorrespondingauthor}}	
\cortext[mycorrespondingauthor]{Corresponding author}
\ead{xie\_wei@scu.edu.cn}

\address[mymainaddress]{Department of Architecture and Civil Engineering\\City University of Hong Kong, Hong Kong}
\address[2]{Department of Construction Management\\Tsinghua University, Beijing, China}
\address[3]{Sichuan University-The Hong Kong Polytechnic University Institute for Disaster Management and Reconstruction\\Sichuan University, Chengdu, China}

\begin{abstract}
Crowd movement simulation is crucial for pedestrian safety management and facility design. Data-driven models offer the potential to improve realism and predictive accuracy, but most are developed for a single scenario, limiting their flexibility. We propose a data-driven crowd simulation model that incorporates refined visual-information extraction and explicit exit cues, aiming to improve flexibility across multiple scenarios by more effectively capturing core navigational features. The model is tested on four fundamental modules (bottleneck, corridor, corner, and T-junction) and further evaluated in a composite scenario using a modular approach. Results show that our model performs well across these scenarios, aligning with pedestrian movement in real-world experiments, and outperforms the classical knowledge-driven model in these scenarios. The research outcomes can provide inspiration for the development of data-driven crowd simulation models and advance the application of data-driven approaches.

\end{abstract}


\end{frontmatter}

\section{Introduction}

Crowd simulation is an essential tool for pedestrian safety management and facility design. The most widely utilized approaches are knowledge-driven models, such as the social force (SF) model \cite{Helbing2000}, the cellular automaton model \cite{VARAS2007631}, and the velocity model \cite{Velocity_Obstacles, 4543489}. These models have achieved notable outcomes, including the simulation of pedestrian movement across diverse scenarios \cite{Steffen2009, MA20102101, ZENG2014143, ZENG201737}, the modeling of various pedestrian behaviors \cite{XIE2021105029, XIE2022105875, Lee2016}, and the reproduction of characteristic self-organization phenomena \cite{helbing2002simulation}. Despite these successes, knowledge-driven models often fall short in realistically representing individual trajectories and velocities \cite{Song2021}, as their governing equations and rules are often too simplistic to capture the complex mechanisms of pedestrian movement \cite{Mayi2019}. \par

Alternatively, deep learning-based data-driven approaches are gaining increasing attention in crowd simulation. Similar to knowledge-driven models, they are employed to simulate pedestrian motion at both macroscopic (e.g., collective flow patterns) \cite{HAN2025114644, LU2025113013, 10843990} and microscopic (e.g., individual trajectories and behaviors) levels. At the microscopic level, such models have been applied to simulate crowd movement in various scenarios, including crosswalks \cite{Mayi2016, Mayi2019}, corridors \cite{ZHAO2020, ZHAO2021, Liang2024, Song2021}, bottlenecks \cite{Song2021, 10077452}, T-junctions \cite{9898931, Liang2024, WANG2025130775}, corners \cite{Liang2024, JIANG2025125706}, and slopes \cite{10808163}. It has been demonstrated that data-driven crowd simulation models achieve more realistic and accurate results than knowledge-driven models, as evidenced by the fundamental diagram and individual movement trajectories \cite{Song2021}. This advantage stems from their ability to learn directly from real-world pedestrian motion data \cite{Mayi2019} and their capacity to capture complex crowd mechanisms through numerous parameters \cite{Liang2024}. \par

Nevertheless, a key limitation persists: most microscopic data-driven models are developed and validated for only a single geometric configuration, severely restricting their adaptability to other layouts. For instance, the model in \cite{ZHAO2021} is designed specifically for corridors, and its performance in other typical geometries (e.g., corners, T-junctions) is unproven. Although some models, like the deep convolutional LSTM network (DCLN), have been applied to two scenarios (rooms and corridors) \cite{Song2021}, those capable of handling more than one scenario remain scarce. This scarcity stems from the fundamental challenge of identifying the core, common information that pedestrians utilize across diverse scenarios. To address this, our previous work \cite{Liang2024} introduced a Visual-Information-Driven (VID) model, inspired by a vision-based flocking theory \cite{vision_flocking}. To our knowledge, VID was the first data-driven model demonstrated to be adaptable across three geometries: corridor, corner, and T-junction. However, the original VID model is not applicable to bottleneck scenarios, as it poses challenges for pedestrians in accurately locating the exit within this framework. \par

Bottlenecks, corridors, corners and T-junctions are often considered the most prevalent modules of walking spaces. Achieving realistic pedestrian movement within them through data‑driven simulation would enable the assembly of complex scenarios—akin to combining building blocks. Particularly, bottlenecks significantly influence passage efficiency and are among the most hazardous areas, making accurate prediction in such scenarios paramount. While the original VID model is unsuitable for bottlenecks, we found it can be extended to handle them by refining its visual‑information extraction and incorporating explicit exit cues. In this work, we therefore propose an Improved VID (IVID) model, which can adapt to four core scenarios: corridors, corners, T‑junctions, and bottlenecks. Furthermore, we introduce a modular approach for applying IVID in composite scenarios. \par

The rest of this paper is organized as follows. Section \ref{Sec:Methodology} introduces the IVID model and the modular approach. Section \ref{Sec:Experiments} details the experiments and simulation results across the four modules. Section \ref{Sec:Modular_application} illustrates the application of the modular approach. Section \ref{Sec:Discussion} discusses the findings, and Section \ref{Sec:Conclusion} presents the conclusions.

\section{Methodology} \label{Sec:Methodology} 
In this section, we present the IVID model, including its overall framework and providing a detailed description of each component. Furthermore, we describe the modular approach for implementing the IVID model in complex scenarios.
\subsection{Model} 
\subsubsection{Model framework}
\begin{figure*}[h]
    \centering
    \includegraphics[width=0.9\textwidth]{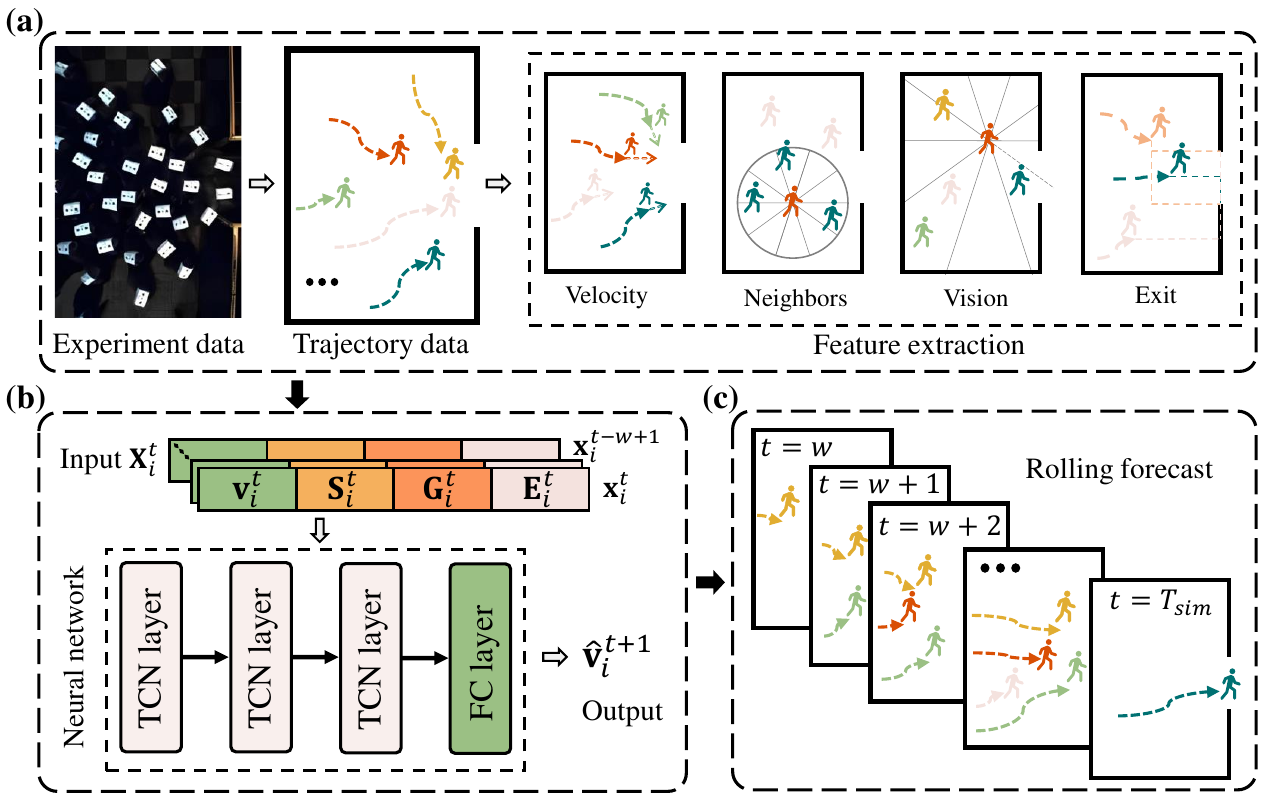}
    \caption{Overall framework of the improved visual-information-driven model. (a) Feature extraction. (b) Velocity-prediction neural network. (c) Rolling forecast.}
    \label{fig:ivid_overview}
    \end{figure*}

The objective of our IVID model is to simulate continuous crowd movement based on the initial motion status of pedestrians. The overall framework of the model is illustrated in Fig. \ref{fig:ivid_overview}. The IVID model comprises three key components: feature extraction, velocity-prediction neural network (VPNN), and rolling forecast. First, the feature extraction component obtains features that influence pedestrian movements from experimental trajectory data, as depicted in Fig. \ref{fig:ivid_overview} (a). Subsequently, these features are input into the VPNN, which outputs the predicted velocities of pedestrians at the next time step, as shown in Fig. \ref{fig:ivid_overview} (b). The VPNN is trained based on the predicted and the ground truth values of velocities. Last, upon completing the training, the rolling forecast component enables continuous crowd movement simulation based on the initial motion status of each pedestrian by iteratively predicting velocity using the trained model and updating pedestrians' positions, as represented in Fig. \ref{fig:ivid_overview} (c).

\subsubsection{Feature extraction} \label{sec:feature_extraction}
The feature extraction component is responsible for extracting features influencing pedestrians' movement. Effectively capturing these features and inputting them into the neural network is paramount for the accuracy of trajectory prediction. Pedestrian movement is primarily influenced by individual velocity, social interactions with neighboring pedestrians, and the physical environment \cite{Helbing2000}. Specifically, the physical environment encompasses visual information and exit information. Visual information provides pedestrians with the geometry of their surrounding walkable space and their own location, thereby facilitating navigation. Exit information indicates the intended destination of the pedestrians. Consequently, the component extracts these influential features, which are detailed below.

\begin{enumerate}
\item Individual velocity ${\mathbf{v}_i^t}$. A pedestrian's future velocity is influenced by their previous velocities due to the effects of inertia \cite{Courtine2003, MA20102160, Yanagisawa_2009}. This correlation justifies the inclusion of individual velocity ${\mathbf{v}_i^t}$ as input for the VPNN. Specifically, ${\mathbf{v}_i^t} \in \mathbb{R}^{1 \times 2}$ represents the magnitude of the velocity components of subject pedestrian $i$ in the $x$- and $y$-axis directions at time step $t$.

\item Social information ${\mathbf{S}_i^t}$. Social information primarily originates from neighboring pedestrians. As pedestrians walk, they exhibit various social behaviors, including following, avoidance, and other interactions with their neighbors \cite{RevModPhys.73.1067, XIE2022104100}. Pedestrians' movements are significantly influenced by their neighboring pedestrians. Several methods have been proposed to identify these neighbors \cite{MA20102101, wkas2006social, Mayi2016}. In our IVID model, as in the original VID model \cite{Liang2024}, we continue to utilize the radar-nearest-neighbor (Radar-NN) method proposed by \cite{ZHAO2021} due to its demonstrated effectiveness. Fig. \ref{fig:Radar-NN_Vision-IVID} (a) illustrates the interaction pattern of the Radar-NN. First, for the subject pedestrian $i$ at time step $t$ (denoted as ($i, t$)), a circular social interaction area is defined, centered at the current position of pedestrian $i$ with an interaction radius $R$. This circular social interaction area is subsequently divided into equally sized subareas $\{a_{i^1}^t, a_{i^2}^t, \cdots, a_{i^j}^t, \cdots, a_{i^{N_j}}^t\}$ beginning from the $x$-axis with a unified central angle $\alpha$. Here, $j$ represents the index, and $N_j=360^\circ/\alpha$ is the number of the subareas. Radar-NN identifies a single nearest neighbor \( nei_j \) for each subarea \( a_{i^j}^t \). Specifically, neighbor \( nei_j \) is defined as
\[
nei_j = 
\begin{cases} 
\arg\min_{m \in \mathcal{N}} d(i, m) & \text{if } \mathcal{N} \neq \emptyset \\ 
\text{midpoint}(a_{i^j}^t) & \text{if } \mathcal{N} = \emptyset 
\end{cases}
\]
where \( \mathcal{N} = \{ m \mid m \text{ is a pedestrian or wall point in } a_{i^j}^t \} \) denotes the set of other pedestrians or wall points within the subarea \( a_{i^j}^t \); \( d(i, m) \) represents the distance between pedestrian \( i \) and entity \( m \); and \( \text{midpoint}(a_{i^j}^t) \) represents the midpoint of the arc corresponding to the sector \( a_{i^j}^t \).
\par

Thus, a total of $N_j$ neighbors are identified. Last, the input feature ${\mathbf{S}_i^t} \in \mathbb{R}^{{N_j} \times 4}$ represents the relative positions and velocities of the $N_j$ neighbors with respect to pedestrian $i$ in the $x$- and $y$-axis directions at time step $t$.

\item Visual information $\mathbf{G}^{t}_i$. Pedestrians inherently perceive their surroundings through visual information, which is essential for navigating various scenarios as it reveals the geometry of walkable spaces and their own location. Consequently, effectively extracting and incorporating visual information is vital for enhancing model flexibility. In the original VID model \cite{Liang2024}, we employ a velocity-direction-based half-vision mode to extract visual information. However, this mode is not adaptable for scenarios that involve frequent backward movement, such as bottleneck situations, where pedestrians are unable to detect the exit while moving backward. Therefore, in the IVID model, we adopt a $x$-axis-based full-vision extraction mode to extend its applicability for bottleneck scenarios. Fig. \ref{fig:Radar-NN_Vision-IVID} (b) illustrates the visual information extraction pattern. Specifically, for ($i, t$), the subject pedestrian $i$ serves as the center, emitting rays $\{r_{i^1}^t, r_{i^2}^t, \cdots, r_{i^k}^t, \cdots, r_{i^{N_k}}^t\}$ originating from the $x$-axis within a 360-degree area at uniform intervals of $\beta$. Here, $k$ represents the index, and ${N_k}=360^\circ/\beta$ is the number of the rays. Each ray $r_{i^k}^t$ corresponds to a unique visual point $k$. Specifically, point $k$ is defined as the nearest intersection point where ray $r_{i^k}^t$ intersects with the scenario wall. If no intersection occurs, $r_{i^k}^t$ points toward the exit. Given that the space in the exit direction is unbounded, we introduce a virtual exit-distance parameter $D_e$, where $D_e$ is a large constant. In this context, point $k$ is defined as the location along the ray $r_{i^k}^t$ that is situated at a distance $D_e$ from pedestrian $i$. Consequently, a total of $N_k$ visual points are identified. The input feature ${\mathbf{G}_i^t} \in \mathbb{R}^{{N_k} \times 2}$ represents the relative positions of the $N_k$ visual points with respect to pedestrian $i$ in the $x$- and $y$-axis directions at time step $t$.

    \begin{figure*}[htpb]
    \centering
    \includegraphics[width=0.9\textwidth]{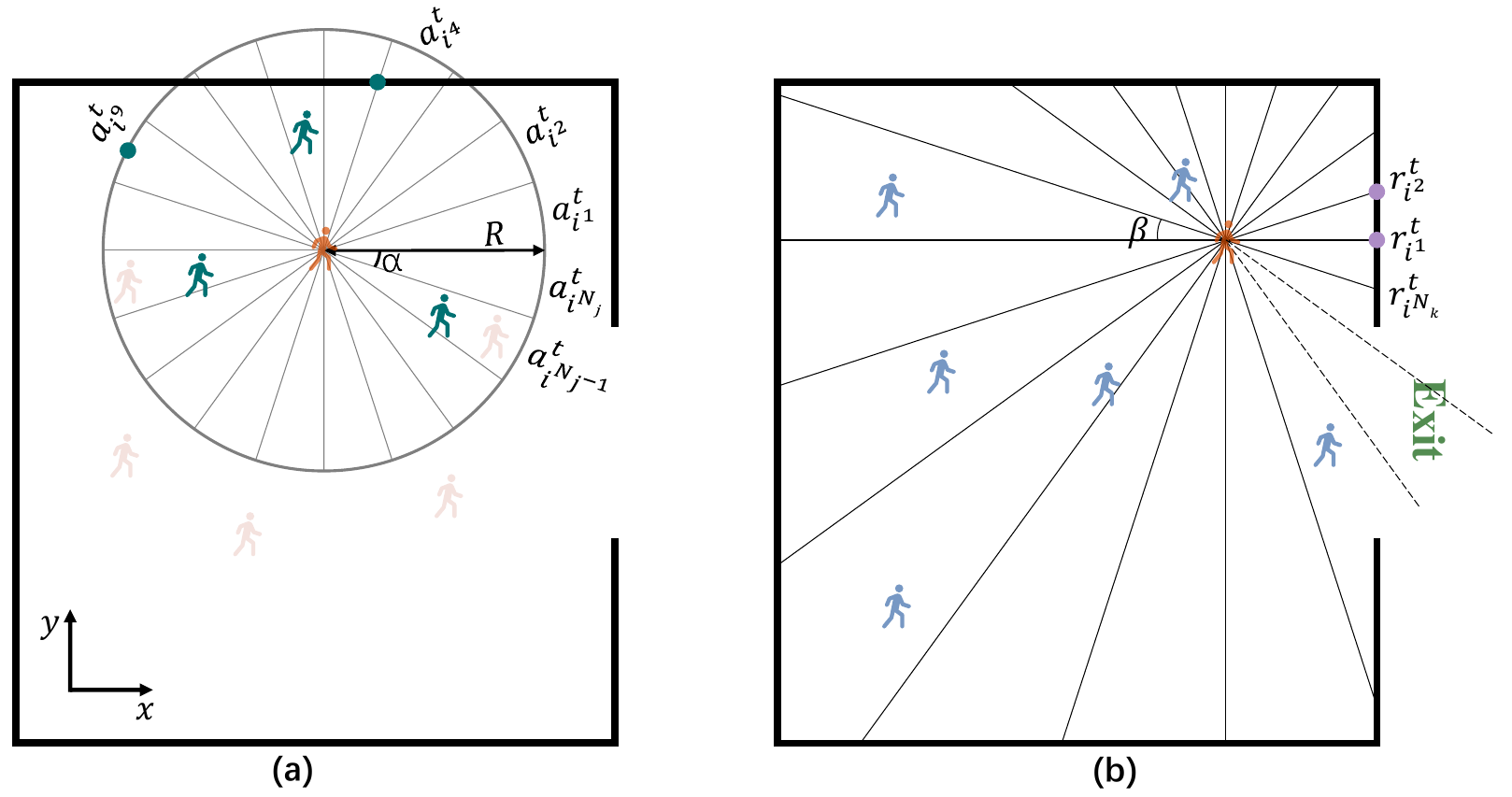}
    \caption{Extraction of social and visual information. (a) Interaction pattern of Radar-NN. The orange pedestrian icon represents the current pedestrian \( i \). The green and pink pedestrian icons denote neighbors and non-neighbors, respectively, within the Radar-NN pattern. If the nearest entity in the subarea \( a_{i^j}^t \) is another pedestrian, that pedestrian is designated as neighbor \( nei_j  \) (green pedestrian icon in \( a_{i^{N_j-1}}^t \)). Conversely, if the nearest entity in the subarea \( a_{i^j}^t \) is a wall point, this wall point is considered neighbor \( nei_j  \) (green point in \( a_{i^4}^t \)). If \( a_{i^j}^t \) contains neither pedestrians nor wall points, the midpoint of the arc corresponding to the sector \( a_{i^j}^t \) is regarded as \( nei_j  \) (green point in \( a_{i^9}^t \)). (b) Visual information extraction pattern. The representation of $r_{i^k}^t$ is indicated by solid and dotted black lines when it intersects with a wall and an exit, respectively. The lengths of the dashed lines are $D_e$. The purple points on $r_{i^1}^t$ and $r_{i^2}^t$ represent the corresponding visual points of $r_{i^1}^t$ and $r_{i^2}^t$.}
    \label{fig:Radar-NN_Vision-IVID}
    \end{figure*}

\item Exit information ${\mathbf{E}_i^t}$. Exit information indicates the intended destinations of pedestrians. Particularly, in bottleneck scenarios, the passage narrows abruptly at the exit, significantly influencing pedestrian movement. Accurate identification of the exit enables pedestrians to effectively plan their routes and navigate the confined space. Therefore, we incorporate exit information into the IVID model, relative to the original VID model \cite{Liang2024}, to enhance its flexibility. Specifically, ${\mathbf{E}_i^t} \in \mathbb{R}^{2 \times 2}$ denotes the relative positions of the two endpoints of the exit with respect to pedestrian $i$ along the $x$- and $y$-axis directions at time step $t$.
\end{enumerate}

Lastly, we reshape ${\mathbf{v}_i^t}$, ${\mathbf{S}_i^t}$, $\mathbf{G}^{t}_i$, and ${\mathbf{E}_i^t}$  into one-dimensional (1D) vectors, and then concatenate these reshaped vectors to $\mathbf{x}_i^t  \in \mathbb{R}^{1 \times (2+4N_j+2N_k+4)}$. 

\subsubsection{Velocity-prediction neural network (VPNN)}

\begin{figure*}[htpb]
    \centering
    \includegraphics[width=0.7\textwidth]{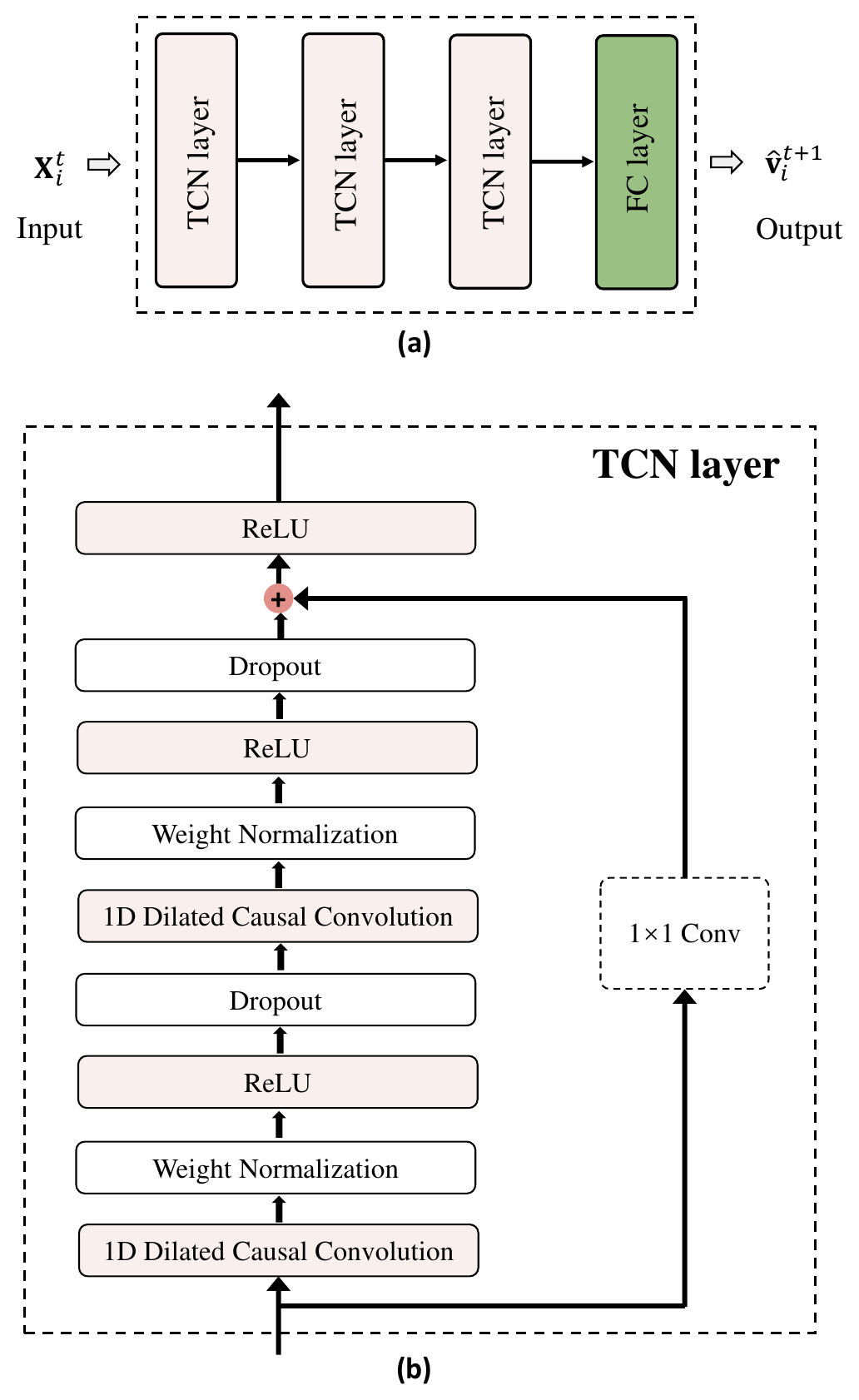}
    \caption{Schematic of the velocity-prediction neural network. (a) Architecture of the velocity-prediction neural network. (b) Architecture of the TCN layer \cite{bai2018empirical}.}
    \label{fig:VPNN}
    \end{figure*}

The VPNN is responsible for predicting pedestrian's velocity at the next time step based on the extracted features from previous time steps. A TCN-based neural network \cite{bai2018empirical, Liang2024} is employed to address this sequence modeling problem. The architecture of the VPNN is depicted in Fig. \ref{fig:VPNN} (a). Specifically, VPNN takes $\mathbf{X}_i^t = [\mathbf{x}_i^{t-w+1}; \mathbf{x}_i^{t-w+2}; \cdots; \mathbf{x}_i^{t}]$ as input and outputs the predicted velocity vector ${\hat{\mathbf{v}}_i^{t+1}}$. Here, $w$ denotes the lookback window, and ${\hat{\mathbf{v}}_i^{t+1}} = [{\hat{v}_{i_x}^{t+1}}, {\hat{v}_{i_y}^{t+1}}] \in \mathbb{R}^{1 \times 2}$. ${\hat{v}_{i_x}^{t+1}}$ and ${\hat{v}_{i_y}^{t+1}}$ denote the predicted velocity components along the $x$- and $y$-axes, respectively. Therefore, VPNN simultaneously predicts both components of the velocity vector. VPNN comprises three TCN layers and one fully connected (FC) layer. The architecture of the TCN layer is illustrated in Fig. \ref{fig:VPNN} (b). It includes 1D dilated causal convolutions, weight normalization \cite{salimans2016weight}, rectified linear unit (ReLU) activation \cite{nair2010rectified}, and dropout \cite{srivastava2014dropout}. Notably, the 1D dilated causal convolution is the most critical operation within the TCN layer, as it ensures that there is no information leakage from future to past and that the input and output sequences of the TCN layer maintain the same length. For a sequence input denoted as $\bm{Z}$ and a filter $f$, the 1D dilated causal convolution operation $\mathbf{\Theta}$ on element $e$ of the sequence is defined as

\begin{equation}
    \mathbf{\Theta}(e) = \sum_{u=0}^{q-1}f(u)\cdot \bm{Z}_{e-h \cdot u}
\end{equation}
where $q$ is the kernel size, $h$ is the dilation factor,  and $u$ is the index over the convolution kernel elements (i.e., $f(u)$ represents the $u$-th element of the kernel $f$). 

\par
We employ the Mean Squared Error (MSE) loss to train the VPNN. For ($i, t$), the loss is defined as 
\begin{equation}
    loss_i^t = ({\hat{v}_{i_x}^{t+1}}-{v}_{i_x}^{t+1})^2 + ({\hat{v}_{i_y}^{t+1}}-{v}_{i_y}^{t+1})^2
\end{equation}
where ${v}_{i_x}^{t+1}$ and ${v}_{i_y}^{t+1}$ denote the ground truth velocity components along the $x$- and $y$-axes, respectively.

\subsubsection{Rolling forecast}
The rolling forecast component is responsible for simulating continuous crowd movement following the completion of VPNN training. The configuration of the rolling forecast simulation scenarios is based on the controlled-experiment scenarios (i.e., the test datasets presented in Table \ref{tab:datasets}), thereby ensuring the comparability of experimental and simulated data. Further information regarding these scenarios is provided in Section \ref{sec:datasets}. The entry times of pedestrians into the scenario and the trajectories of their initial $w$ time steps are consistent with those in the corresponding controlled experiments. The simulation for each pedestrian commences once the length of their previous trajectory sequence reaches $w$. 
\par
At each time step $t$, the influencing feature $\mathbf{X}_i^t$ of each pedestrian $i$ is extracted and subsequently input into the trained VPNN model to predict the velocity at the next time step, ${\hat{\mathbf{v}}_i^{t+1}}$. Based on the current position of each pedestrian $i$ and the predicted velocity ${\hat{\mathbf{v}}_i^{t+1}}$, the new position of each pedestrian $i$ can be updated. This process is repeated until all pedestrians have exited the scenario.

\subsection{Modular approach} \label{sec:modular_approach}
We propose a modular approach for applying the IVID model in complex scenarios. Various complex scenarios are constructed from simple scene modules, the most common of which include corridors, bottlenecks, corners, and T-junctions. Leveraging the IVID model's effective performance across these four fundamental modules, we can assemble them into diverse complex scenarios, similar to building with Lego, thereby facilitating the simulation of intricate environments.  \par
\begin{figure*}[htpb]
    \centering
    \includegraphics[width=0.9\textwidth]{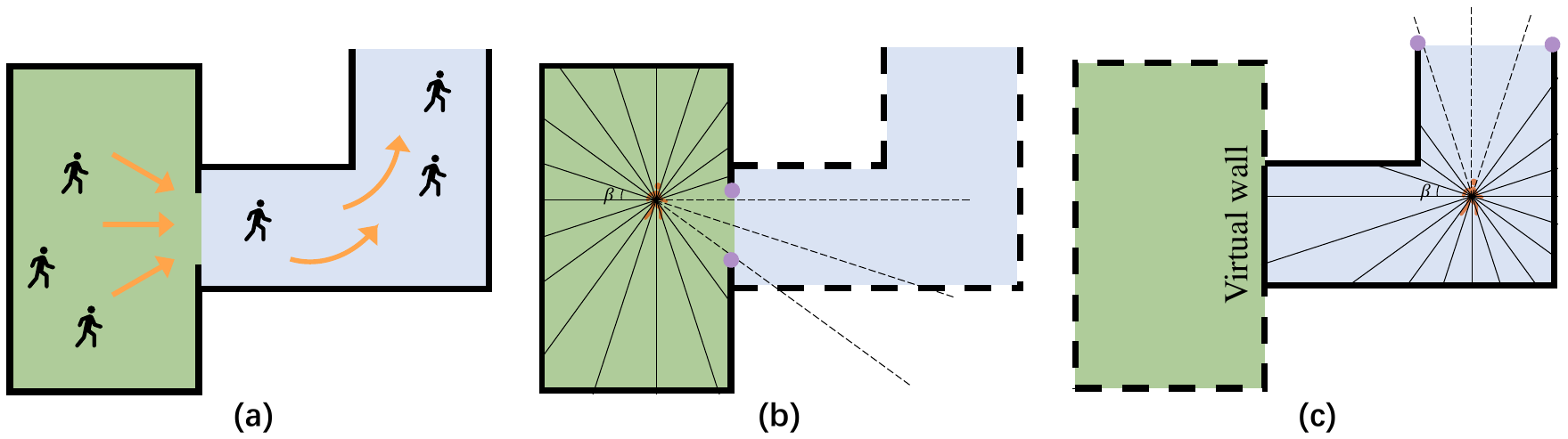}
    \caption{Schematic of the modular approach. (a) An example scenario comprising a bottleneck and a corner, with the bottleneck module represented by the green area and the corner module represented by the blue area. (b) Schematic illustrating the extraction of visual and exit information for a pedestrian located within the bottleneck module. (c) Schematic illustrating the extraction of visual and exit information for a pedestrian situated within the corner module.}
    \label{fig:modular_approach}
    \end{figure*}

The schematic representation of the modular approach is illustrated in Fig. \ref{fig:modular_approach}. The core principle of this approach is that a pedestrian, when situated within a specific module, considers only the influences of walls within that same module, effectively disregarding interactions from walls in other modules. Furthermore, the target exit for each pedestrian corresponds to the exit of the module in which they are located. For instance, Fig. \ref{fig:modular_approach} depicts a scenario composed of a bottleneck and a corner. The orange arrows indicate the direction of pedestrian movement. When pedestrian $i$ is situated within the bottleneck module, as shown in Fig. \ref{fig:modular_approach} (b), only the walls of the bottleneck, represented by solid lines, are considered for the extraction of visual information $\mathbf{G}^{t}_i$. Additionally, during the extraction of exit information ${\mathbf{E}_i^t}$, the exit points of the bottleneck, indicated as purple points in Fig. \ref{fig:modular_approach} (b), are identified. When pedestrian $i$ exits the bottleneck and is located within the corner module, the extraction of visual information and exit information is illustrated in Fig. \ref{fig:modular_approach} (c). In this context, the walls of the bottleneck are disregarded, while the walls of the corner are taken into account for the extraction of visual information $\mathbf{G}^{t}_i$. Furthermore, a virtual wall is introduced at the junction between the bottleneck and the corner. Simultaneously, the exit information ${\mathbf{E}_i^t}$ is identified as the exit of the corner, represented as purple points in Fig. \ref{fig:modular_approach} (c).

\section{Experiments and results} \label{Sec:Experiments}
\subsection{Experiments}
\subsubsection{Datasets} \label{sec:datasets}
We evaluate the IVID model in four scenarios: bottleneck, corridor, corner and T-junction. The datasets utilized in our study originate from a series of controlled experiments conducted within these four geometries by the Institute of Civil Safety Research at the Research Centre Jülich, Germany. The experimental data are accessible through the pedestrian dynamics data archive (https://ped.fz-juelich.de/da/doku.php). A concise overview of the experiments is provided below.
\par

\begin{figure*}[htpb]
\centering
\includegraphics[width=0.8\textwidth]{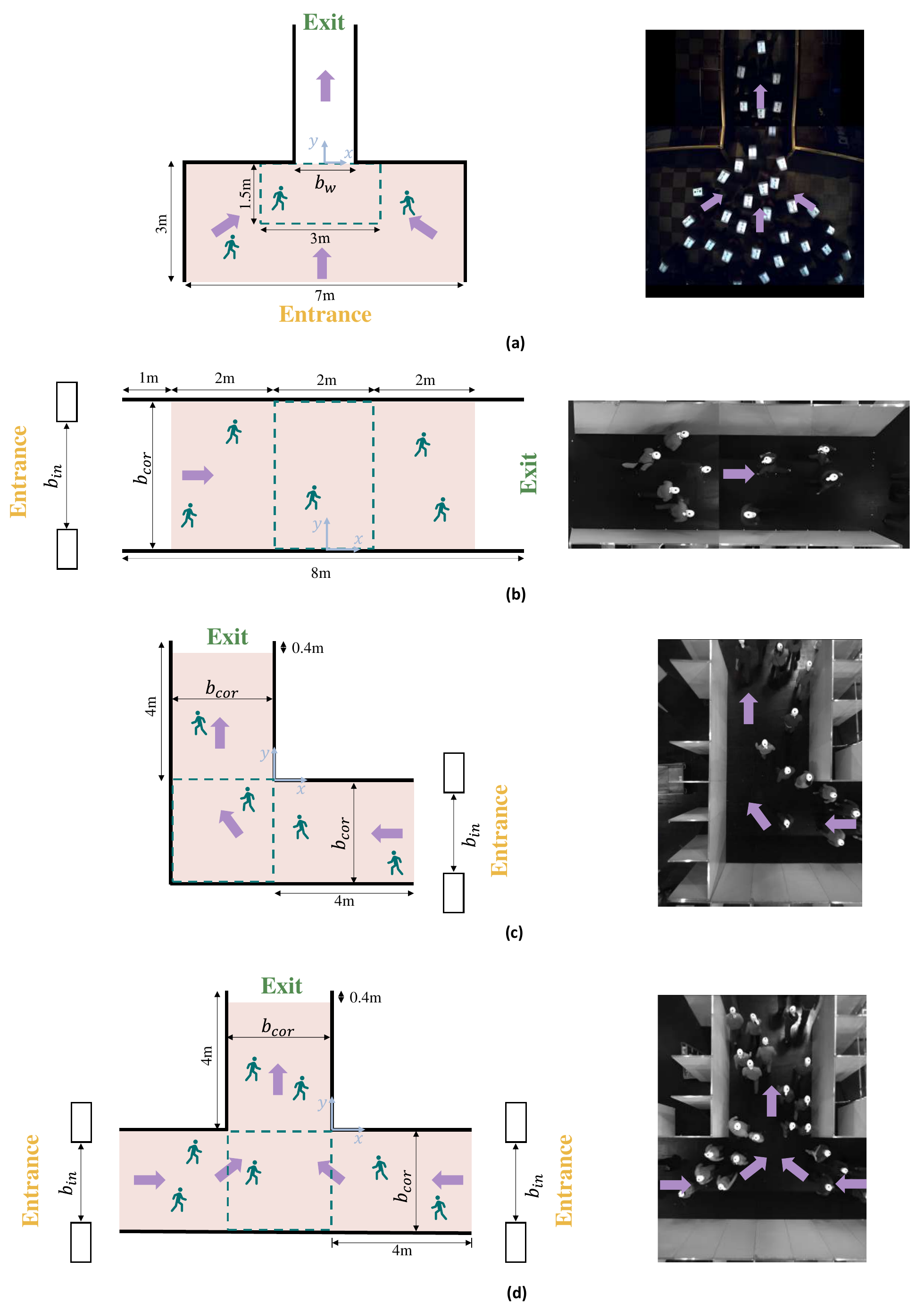}
\caption{Sketches and snapshots of the controlled experiments (https://ped.fz-juelich.de/da/doku.php). (a) bottleneck. (b) corridor. (c) corner. (d) T-junction.}
\label{fig:exp_setup}
\end{figure*}

Fig. \ref{fig:exp_setup} illustrates the experimental setup along with snapshots from the experiments. In each run, pedestrians are instructed to enter the scenario through the designated entrance and exit via the specified exit. The purple arrows in Fig. \ref{fig:exp_setup} indicate the direction of pedestrian movement within each scenario. Multiple runs were conducted for each scenario. Specifically, in bottleneck experiments, the exit width ($b_w$) was varied across different runs. In corridor, corner and T-junction experiments, the width of the entrance ($b_{in}$) and the width of the corridor ($b_{cor}$) were altered in different runs to regulate pedestrian density within the experimental area. Note that the architectural layouts and entrance configurations (as shown in Fig. \ref{fig:exp_setup}) for the bottleneck and T-junction scenarios are different. The bottleneck scenario resembles a process of exiting from a room through an exit. In contrast, the entrances in the T-junction scenario are located on both sides of the corridor, resulting in merging behavior as pedestrian flows converge from either side in the T-junction. The entire experimental process was recorded on video, and pedestrian trajectories were extracted at a rate of 25 fps for the bottleneck scenario and 16 fps for the corridor, corner and T-junction scenarios. We mainly focus on the pink area in Fig. \ref{fig:exp_setup} for each scenario. In other words, we utilize the trajectory data from this region to construct the dataset while simulating pedestrian movement within this area. 
\par
We divided the multiple runs from each scenario into training-validation runs and test runs, as illustrated in Table \ref{tab:datasets}.  We extracted the input features and output targets utilizing the feature extraction component described in Section \ref{sec:feature_extraction} from all the training-validation runs to form the training-validation samples used for model training and validation. All training-validation samples from the four scenarios were pooled together, and these pooled samples were then split into a training set and a validation set at a ratio of 4:1. Specifically, the training set was used to update model parameters during training, while the validation set was employed to select the final model based on the minimum validation MSE loss. After training, we conducted crowd simulations based on the test runs with the rolling forecast component to evaluate the model's performance. The test runs were completely independent of the training-validation runs, ensuring that the test set was never used for any model training or selection decisions.

\begin{table*}[htb]
\centering
\footnotesize
\caption{Training-validation and test runs for each scenario. }
\resizebox{1.02\textwidth}{!}{
\begin{tabular}{ccc}
  \toprule 
  Dataset & Scenario & Name \\ 
  \hline
    \multirow{5}{*}{Training-validation} & Bottleneck                 & W110, W120, W140, W180, W200, W250 \\
                                         \cline{2-3}
                                         & Corridor  & E050-C180,  E060-C180,  E070-C180,  E100-C180,  E145-C180,  E180-C180, E065-C240, E080-C240,  E095-C240,  E145-C240,  E190-C240,  E240-C240\\
                                         \cline{2-3}
                                         & Corner                     & E050-C240,  E060-C240,  E080-C240,  E100-C240,  E150-C240,  E240-C240   \\
                                         \cline{2-3}
                                         & T-junction                 & E050-C240,  E060-C240,  E080-C240,  E100-C240,  E120-C240,  E150-C240,  E240-C240  \\
    \hline
    \multirow{4}{*}{Test}                & Bottleneck                 & W120, W160, W220  \\
                                         \cline{2-3}
                                         & Corridor                   & E080-C300,  E100-C300,  E120-C300,  E180-C300,  E240-C300,  E300-C300  \\
                                         \cline{2-3}
                                         & Corner                     & E050-C300,  E060-C300,  E080-C300,  E100-C300,  E150-C300,  E300-C300   \\
                                         \cline{2-3}
                                         & T-junction                 & E050-C300,  E080-C300,  E120-C300,  E150-C300 \\ 
    \bottomrule
\end{tabular}}
\label{tab:datasets}
\vspace{-1em}
\begin{justify}
    Note: In the bottleneck scenario, the symbols 'W' followed by a width value denote controlled experiments with varying exit widths ($b_w$), where 'W110' indicates an exit width of 110 cm. In the corridor, corner and T-junction scenarios, these symbols represent controlled experiments with different combinations of entrance width ($b_{in}$) and corridor width ($b_{cor}$); for instance, 'E050-C180' signifies an entrance width of 50cm and a corridor width of 180cm. It is important to note that 'W120' in the Training-validation runs and the test runs for the bottleneck scenario refers to two distinct experiments.
\end{justify}
\end{table*}

\subsubsection{Settings}
We set the interaction radius $R$ and the central angle $\alpha$ to 1.2m and 18$^\circ$ , respectively, for social information extraction, based on their demonstrated effectiveness \cite{ZHAO2021, Liang2024}. To evaluate the robustness and efficacy of the full-vision extraction mode, we conduct a sensitivity analysis on the virtual exit-distance parameter $D_e$ and the interval $\beta$. Specifically, $D_e$ is tested at distances of 20m and 100m, while $\beta$ is examined at angles of 5$^\circ$, 10$^\circ$, 15$^\circ$ and 18$^\circ$. Following previous studies on trajectory prediction \cite{9043898, 7780479, DBLP}, we set the lookback window $w$ to 8 time steps. The Adam optimizer \cite{kingma2017adam} is employed to train the VPNN, with a learning rate of 0.0001 and a total of 3000 iterations. The kernel size $q$ is consistently set to 8 for each TCN layer to provide a sufficient receptive field. The dilation factors $h$ for the three TCN layers are configured as 1, 2 and 4, respectively, facilitating exponential dilation. Additionally, the channels for the convolutional layers in the three TCN layers are set to 32, 64 and 96.

\par
Since our model employs a data-driven approach, deviations in pedestrian trajectory predictions may occur, leading to minor occurrences of trajectory passing through walls during the rolling forecast simulation. In such instances, following the methods outlined in \cite{ZHAO2021, Liang2024}, we reset the trajectories of the pedestrians for the preceding $w$ time steps. Specifically, pedestrians are set to move along the walls in corridor, corner and T-junction scenarios, or towards the exit in bottleneck scenarios during these corrected $w$ time steps, with the velocity magnitude set equal to the average speed computed from their original predicted velocities over the same $w$-step window. Subsequently, we re-extract and update the input features for the preceding $w$ time steps of the pedestrians based on the corrected trajectories to predict future velocities in a rolling manner.

\subsection{Results}
\subsubsection{Parameter sensitivity analysis} \label{sec:parameter_senti}
We conducted a sensitivity analysis on the virtual exit-distance parameter $D_e$ and the interval $\beta$ to evaluate the robustness of the proposed full-vision extraction mode. Three quantitative metrics are used to assess the parameter sensitivity based on the trajectories obtained from the controlled experiments and our simulation: 
\begin{enumerate}
\item Average Displacement Error (ADE): The mean Euclidean distance between trajectory points in the simulation and those from the corresponding controlled experiments. 
\item Final Displacement Error (FDE): The Euclidean distance between the final locations in the simulation and the controlled experiments.
\item Travel Time Error (TTE): The difference between the simulated travel time and the actual travel time.
\end{enumerate}

\begin{figure}[htpb]
\centering
\includegraphics[width=0.45\textwidth]{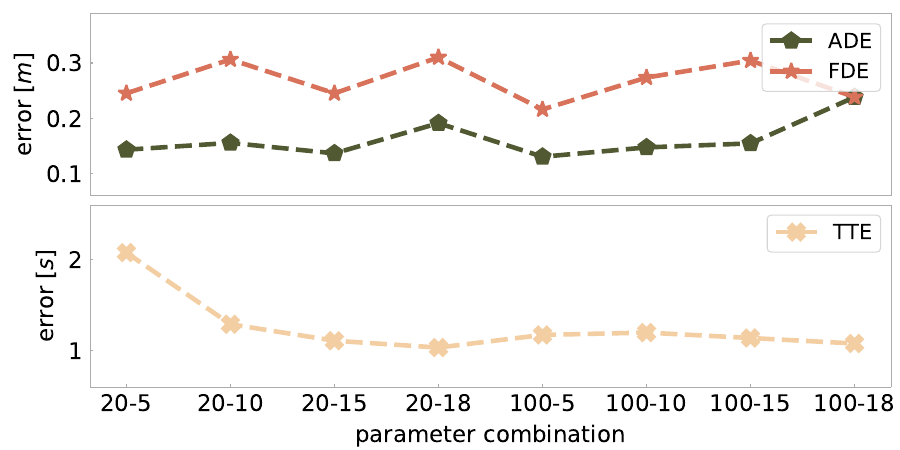}
\caption{Mean ADE, FDE and TTE in various parameter combinations. The x-axis denotes the combinations of $D_e$ and $\beta$.}
\label{fig:param_sens}
\end{figure}

The mean values across all pedestrians of these three metrics for different parameter combinations are presented in Fig. \ref{fig:param_sens}. It can be observed that our model maintains relative stability within the tested parameter ranges of $D_e$ and $\beta$, as the three metric values exhibit minimal variation across different parameter combinations. This indicates the robustness of the proposed full-vision extraction mode and a substantial parameter selection space for $D_e$ and $\beta$.

\subsubsection{Qualitative and quantitative comparisons}
We compare the simulation results of the proposed IVID model and the SF model \cite{Helbing2000}, along with data obtained from the control experiments, to assess the validity of the IVID model. Specifically, we reproduce the SF model in the testing scenarios of the four geometries (i.e., bottleneck, corridor, corner and T-junction), respectively. The direction of the desired velocity in the corner and T-junction scenarios is determined by the method proposed in \cite{mohcine2013}. The magnitude of the desired velocity is set to follow a normal distribution with a mean of 1.4m/s and a variance of 0.2m/s, as the average speed of pedestrians during conditions approaching free movement in controlled experiments is approximately 1.4m/s \cite{Liang2024}. Both qualitative metrics (i.e., trajectories and fundamental diagrams) and quantitative metrics (i.e., ADE, FDE and TTE) are employed to evaluate and compare simulation performance. The results are detailed below. \par

\begin{figure}[htpb]
\centering
\includegraphics[width=0.98\textwidth]{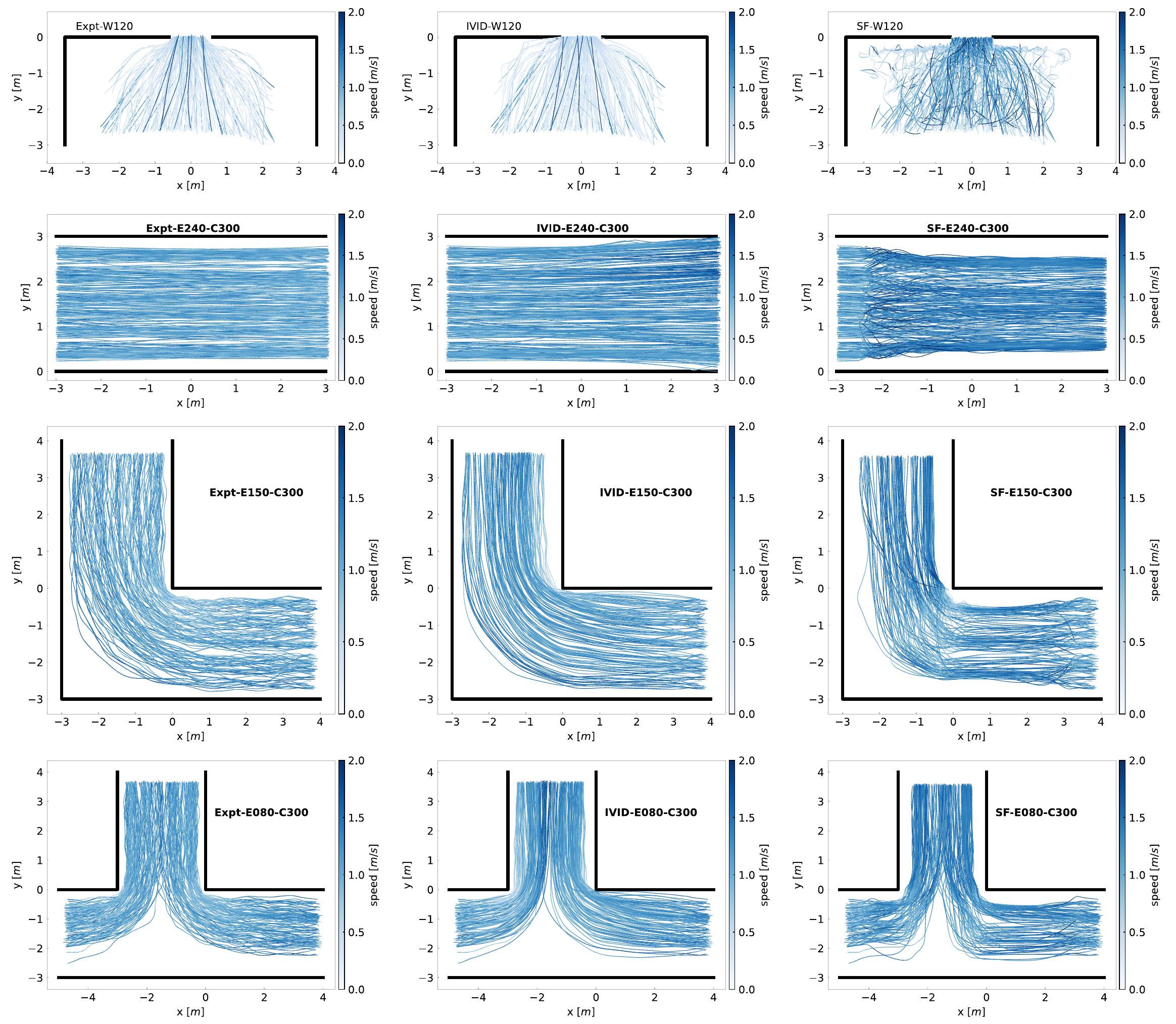}
\caption{Trajectories from the controlled experiments and simulations of the proposed IVID model (with parameters $D_e=100m$ and $\beta=5^\circ$) and the SF model \cite{Helbing2000} in bottleneck, corridor, corner and T-junction scenarios.}
\label{fig:trajs}
\end{figure}

\begin{figure}[htpb]
\centering
\includegraphics[width=0.98\textwidth]{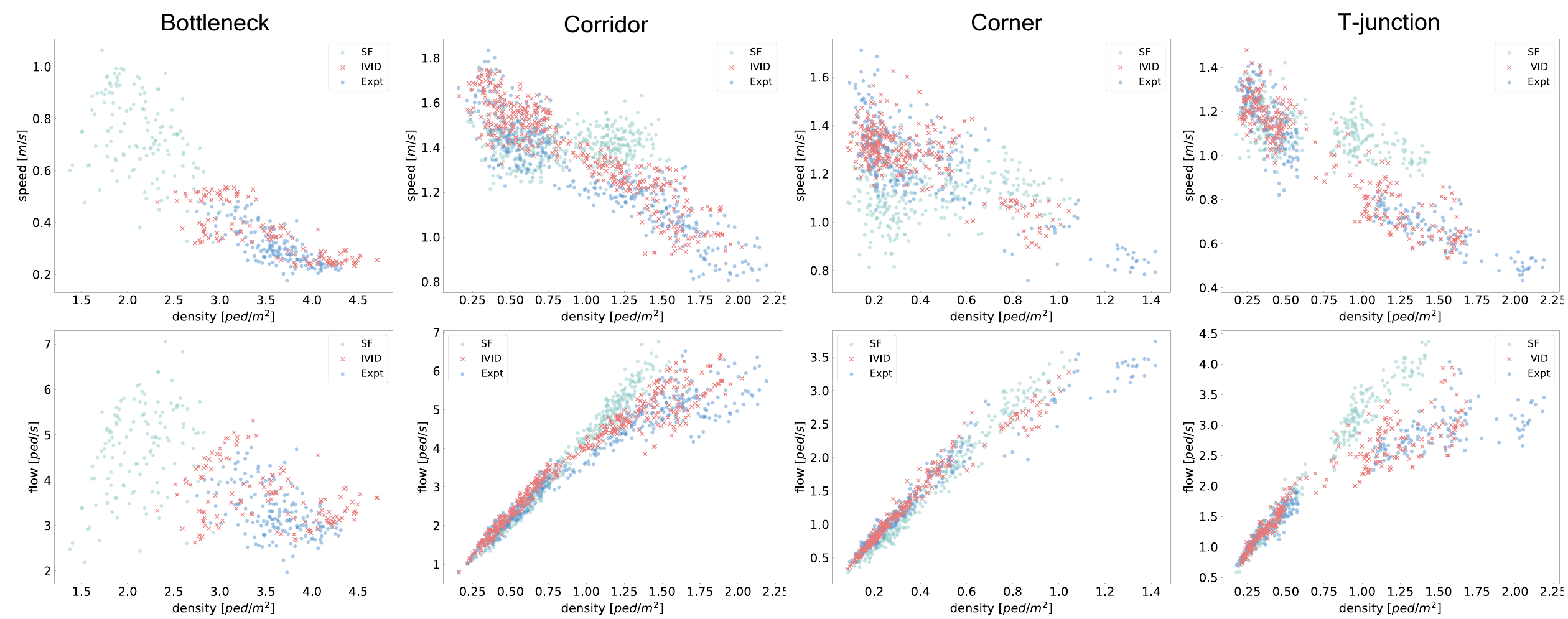}
\caption{Fundamental diagrams derived from controlled experiments, our IVID model (with parameters $D_e=100m$ and $\beta=5^\circ$) and SF model \cite{Helbing2000} for the bottleneck, corridor, corner and T-junction geometries, respectively.}
\label{fig:FD}
\end{figure}

\textbf{Trajectories}. The trajectories from the controlled experiments, as well as simulations of the proposed IVID model and the SF model \cite{Helbing2000}, for bottleneck, corridor, corner and T-junction scenarios are presented in Fig. \ref{fig:trajs}. The IVID trajectories shown in Fig. \ref{fig:trajs} (and the fundamental diagrams in Fig. \ref{fig:FD}) correspond to the parameter combination $D_e=100m$ and $\beta=5^\circ$, as it delivers a relatively balanced and robust performance across the three evaluation metrics (ADE, FDE, and TTE) in Fig \ref{fig:param_sens}. Due to space limitations, one testing scenario for each geometry is included. The trajectory comparison reveals that our IVID model outperforms the SF model in terms of trajectory shape similarity and speed alignment. The trajectories generated by the IVID model closely align with those from the controlled experiments in both trajectory shape and speed magnitude. The pedestrian speeds in the SF model are considerably higher than those observed in both the controlled experiments and the IVID model when the latter two exhibit low speeds.
\par
\textbf{Fundamental diagram}. Fundamental diagrams illustrate the relationship between pedestrian flow, speed and density, serving as a widely used evaluation metric \cite{Zhang_2011}. The measurement areas for each scenario are represented by the green dashed rectangular regions in Fig. \ref{fig:exp_setup}. Fig. \ref{fig:FD} presents the fundamental diagrams derived from the controlled experiments, the proposed IVID model and the SF model \cite{Helbing2000} for bottleneck, corridor, corner and T-junction scenarios. The fundamental diagrams obtained from our model closely resemble those from the controlled experiments and are more aligned with them than those of the SF model, particularly in terms of the density range and the variation of speed with density.

\textbf{ADE, FDE and TTE}. We evaluate the quantitative performance of the proposed IVID model using three metrics: ADE, FDE and TTE, which are introduced in Section \ref{sec:parameter_senti}. The IVID model is systematically compared with the SF model \cite{Helbing2000} and the original VID model \cite{Liang2024} across bottleneck, corridor, corner and T-junction scenarios, as detailed in Table \ref{tab:ADE_FDE_TTE}. Each metric value for the IVID model is calculated from all 4×2=8 parameter combinations for $D_e$ and $\beta$ to ensure a comprehensive comparison and robust evaluation of the model's performance. The results indicate that the proposed IVID model performs well across all four geometries, surpassing the performance of the other two models. Notably, the IVID model significantly outperforms the SF model in ADE and TTE within bottleneck scenarios, while the metric values for the original VID model in bottleneck scenarios are absent due to its inapplicability to this geometry. In corridor, corner and T-junction scenarios, the IVID model and the original VID model demonstrate comparable performance, with the exception that the IVID model exhibits superior FDE in the corner scenario, while the original VID model shows lower TTE in the T-junction scenario. Overall, both the IVID model and the original VID model perform better than the SF model, particularly in terms of TTE.

\begin{table}[htb]
\centering
\scriptsize
\caption{Mean values of ADE, FDE and TTE obtained from the proposed IVID model, the SF \cite{Helbing2000} model and the original VID model \cite{Liang2024} for bottleneck, corridor, corner and T-junction scenarios.}
\begin{tabular}{    
>{\centering}m{0.07\textwidth} 
>{\centering\arraybackslash}m{0.07\textwidth}
>{\centering\arraybackslash}m{0.07\textwidth}
>{\centering\arraybackslash}m{0.07\textwidth}
>{\centering\arraybackslash}m{0.07\textwidth}
>{\centering\arraybackslash}m{0.1\textwidth}}
  \toprule 
  & & Bottleneck & Corridor &Corner & T-junction \\
\hline
\multirow{2}*{ADE$[m]$}
& IVID & 0.13 & 0.09 & 0.21 & 0.19 \\
& SF & 1.17 & 0.16 & 0.26 & 0.22 \\
& VID & / & 0.10 & 0.24 & 0.17\\
\hline
\multirow{2}*{FDE$[m]$}
& IVID & 0.19 & 0.20 & 0.39 & 0.32\\
& SF & 0.19 & 0.25 & 0.42 & 0.33\\
& VID & / & 0.25 & 0.49 & 0.29 \\
\hline
\multirow{2}*{TTE$[s]$}
& IVID & 1.78 & 0.46 & 1.33 & 1.95 \\
& SF & 5.37 & 1.04 & 1.97 & 3.06 \\
& VID & / & 0.47 & 1.08 & 1.23 \\
\bottomrule
\end{tabular}    
\label{tab:ADE_FDE_TTE}
\end{table}  

\subsubsection{Ablation Study}
Compared to the original VID \cite{Liang2024} model, IVID changes the visual information extraction from a half-vision mode to a full-vision mode and incorporates explicit exit information to accommodate bottleneck scenarios. An ablation study is conducted to quantify the individual contributions of each design choice. The results for HE-model (half-vision with exit information), F-model (full-vision without exit information), and the proposed IVID in bottleneck scenarios are reported in Table \ref{tab:ablation_Study}. Overall, IVID, which integrates full-vision mode and explicit exit information, achieves the best performance among the variants. Notably, the HE-model is substantially worse than both the F-model and IVID across all metrics (ADE, FDE, and TTE). In addition, IVID improves upon the F-model, providing gains in terms of FDE and TTE. These findings suggest that, in bottleneck scenarios, the full-vision mode plays a critical role for improving performance, while explicit exit information further refines prediction accuracy. This is because the full-vision mode offers comprehensive visual coverage, allowing for more reliable identification of the exit and the geometry of the bottleneck.

\begin{table}[htb]
    \centering
    \fontsize{5}{6}\selectfont 
    \caption{Ablation study results comparing IVID with its variants on ADE, FDE, and TTE.}
    \resizebox{0.42\textwidth}{!}{
    \begin{tabular}{cccc}
      \toprule 
      Name & ADE$[m]$ & FDE$[m]$ & TTE$[s]$\\
      \hline
      HE-model & 0.20 & 0.44 & 7.01\\
      F-model & 0.13 & 0.28 & 2.05\\
      IVID & 0.13 & 0.19 & 1.78\\
    \bottomrule
    \end{tabular}}
    \label{tab:ablation_Study}
\end{table}

\section{Modular Application}\label{Sec:Modular_application}
\begin{figure}[htpb]
\centering
\includegraphics[width=0.48\textwidth]{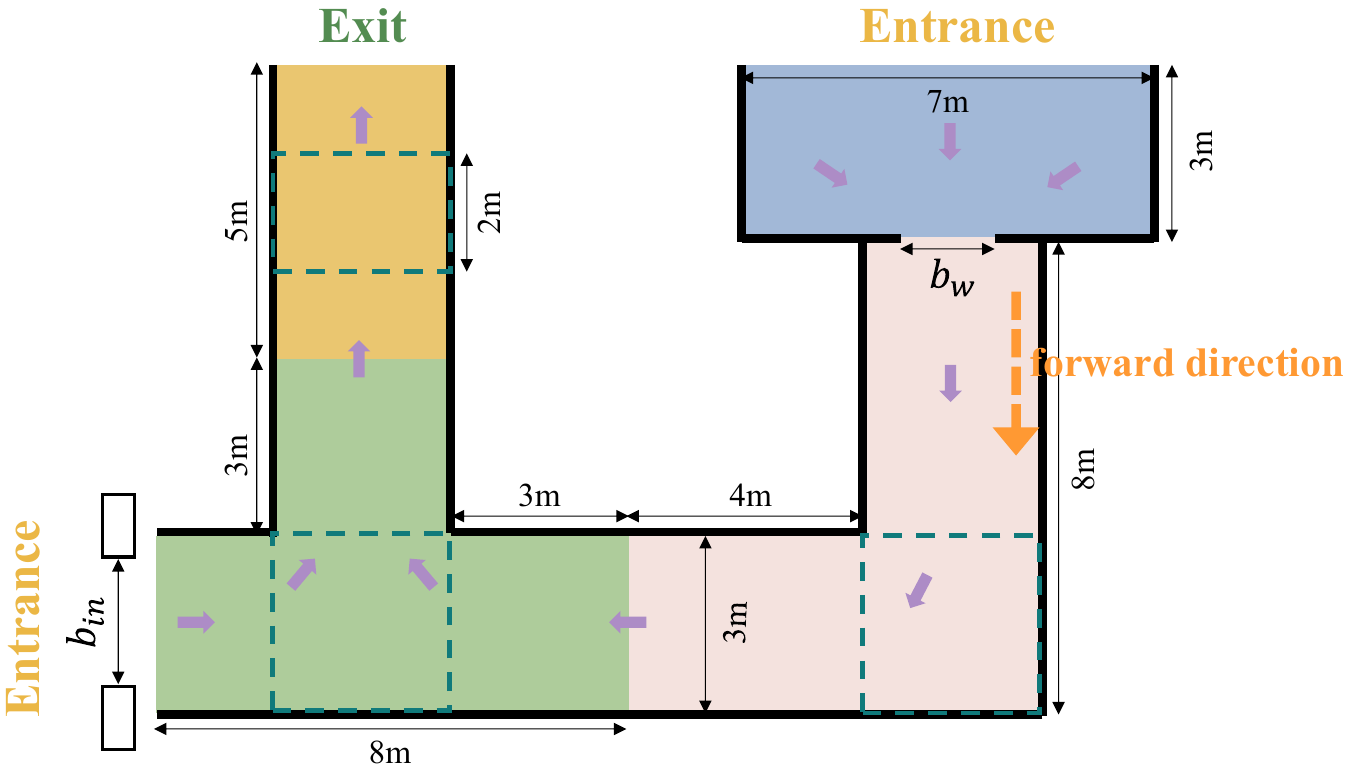}
\caption{Sketch of the composite scenario consisting of a bottleneck, a corner, a T-junction and a corridor module. These four modules are represented by blue, pink, green and yellow regions, respectively.}
\label{fig:complex_scenario_setup}
\end{figure}

We employ the modular approach outlined in Section \ref{sec:modular_approach} to implement the IVID model in composite scenarios. We conduct simulations in a composite scenario utilizing the trained IVID model through modular application to assess model performance. This scenario includes a bottleneck, a corner, a T-junction and a corridor, as illustrated in Fig. \ref{fig:complex_scenario_setup}. Pedestrians enter the scenario from either the bottleneck module or the T-junction module and exit through the corridor module. The purple arrows indicate the direction of movement. A total of four simulation runs were conducted by varying the width of the bottleneck exit ($b_w$) and the entrance width of the T-junction ($b_{in}$), with the combinations summarized in Table \ref{tab:complex_scenarios}. These four $b_w$-$b_{in}$ combinations facilitate a range of pedestrian densities within the simulation, from low to high. To enhance the realism of pedestrian entry into the simulation scenario, the timing of pedestrians entering the bottleneck module and the T-junction module, as well as their initial motion states during the first 8 time steps (lookback window), are consistent with the corresponding controlled experiments. For instance, the initial motion states of pedestrians entering the complex scenario W120-E050 correspond to those from the bottleneck scenario W120 and the T-junction scenario E050-C300 in the controlled experiments. Additionally, we conducted simulations using the SF model \cite{Helbing2000} in the same scenario for model performance comparison. \par

At the boundry between modules, visual information can become inconsistent. For instance, when a pedestrian moves from the bottleneck to the corner at timestep $t$, the visual information from $t-7$ to $t-1$ is derived from the bottleneck, whereas the current visual input at $t$ comes from the corner. This discrepancy renders the use of these eight frames inappropriate for predicting $\mathbf{v}_{t+1}$. Considering this, for time steps $t+1$ to $t+7$, the pedestrian's velocity magnitude in the forward direction (as indicated by the orange dashed arrow in Fig. \ref{fig:complex_scenario_setup}) is set to the average velocity magnitude of the preceding eight time steps. The perpendicular velocity component is set to zero. Using rolling prediction, the pedestrian’s positions for time steps $t+1$ to $t+7$ are updated based on these velocity assignments and the previous timestep’s position. Given that 8 timesteps span only 0.5 seconds (at 16 fps), the resulting simulated trajectory segment is short. This procedure ensures that we obtain a continuous 8-timestep lookback window of velocity and position data after the pedestrian enters the corner, and this $8$-timestep continuous data allows us to effectively implement our subsequent rolling forecast via the trained model. Consequently, after timestep $t+7$, the pedestrian's trajectory is predicted in a rolling manner using our trained model.

\begin{figure}[htpb]
\centering
\includegraphics[width=0.98\textwidth]{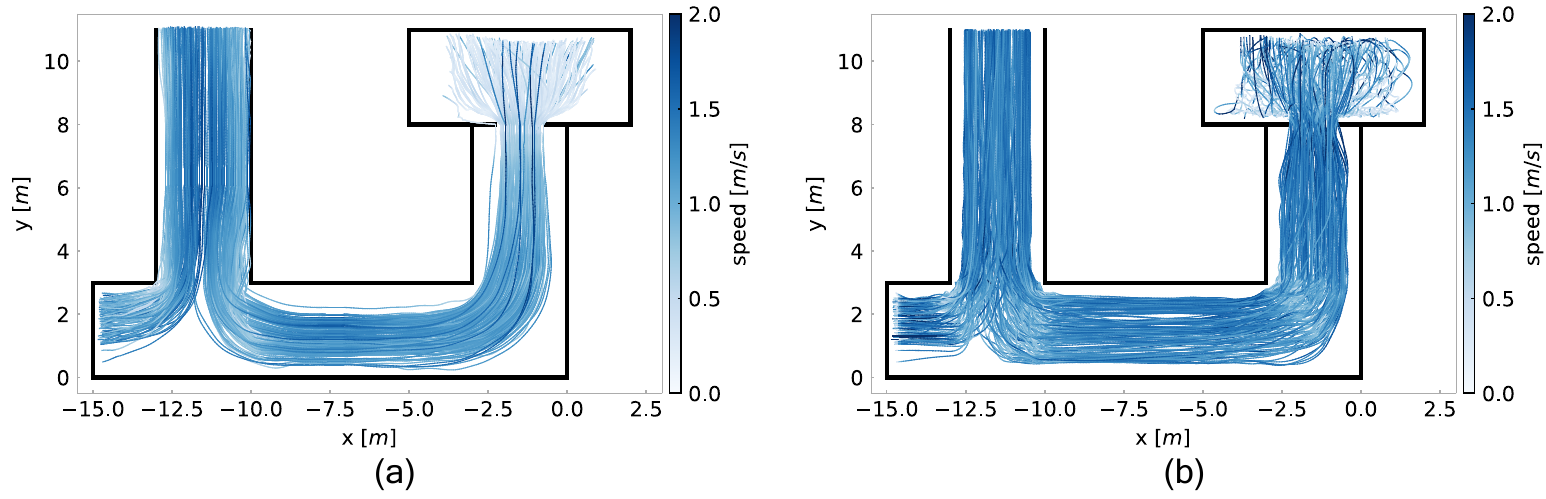}
\caption{Simulation trajectories for scenario W160-E080: (a) from our IVID model with parameters $D_e=100m$ and $\beta=5^\circ$ and (b) from the SF model \cite{Helbing2000}.}
\label{fig:Traj_complex}
\end{figure}

\begin{figure}[htpb]
\centering
\includegraphics[width=0.98\textwidth]{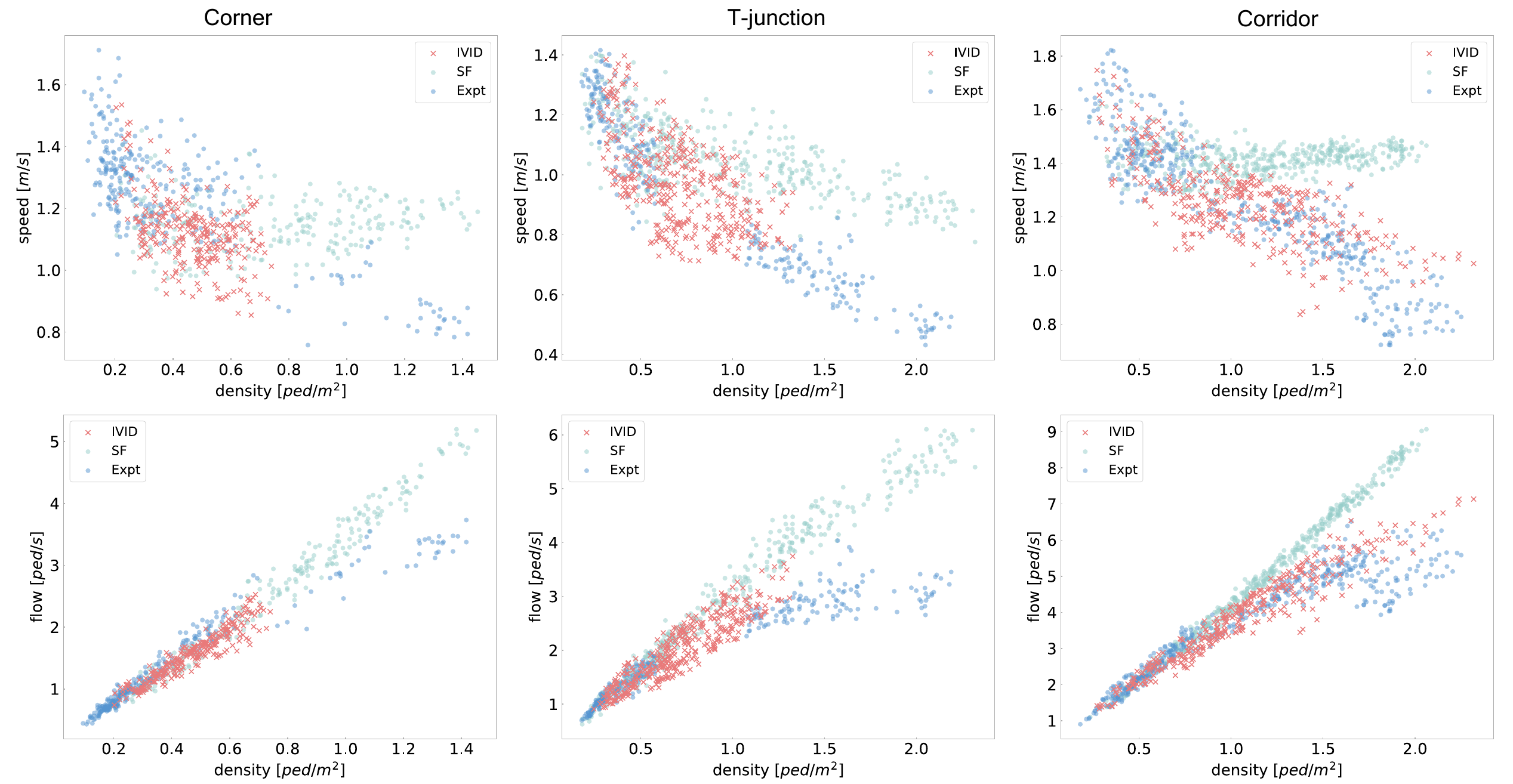}
\caption{Fundamental diagrams for the corner, T-junction and corridor modules obtained from controlled experiments, our IVID model and the SF model \cite{Helbing2000}.}
\label{fig:FD_complex}
\end{figure}

Fig. \ref{fig:Traj_complex} presents the simulation trajectories for scenario W160-E080, utilizing both our IVID model and the SF model \cite{Helbing2000}. The simulation trajectories from our IVID model align well with real-world patterns, characterized by reasonable trajectory shapes and variations in speed. Specifically, as pedestrians exit the bottleneck and enter the corner, their speed gradually increases to a stable level due to the widening of the corridor. The speed of pedestrians at the turn is lower than that in the straight corridor. These observations indicate that our model effectively captures the mechanisms underlying changes in pedestrian velocity. To further evaluate performance, we calculate the fundamental diagrams from these simulated trajectories. Specifically, we compute and compare the fundamental diagrams from the corner, T-junction and corridor modules with those obtained from controlled experiments in these geometries. The measurement areas are represented by the green dashed boxes in Fig. \ref{fig:complex_scenario_setup}. As shown in Fig. \ref{fig:FD_complex}, our IVID model, through the modular approach, effectively reproduces the empirical density–speed–flow relationships in the composite scenario, outperforming the SF model.

\begin{table}[htb]
    \centering
    \fontsize{5}{6}\selectfont 
    \caption{Simulation sceanrios with varying $b_w$ and $b_{in}$.}
    \resizebox{0.42\textwidth}{!}{
    \begin{tabular}{ccc}
      \toprule 
      Name & $b_w[cm]$ & $b_{in}[cm]$\\
      \hline
      W120-E050 & 120 & 50 \\
      W160-E080 & 160 & 80 \\
      W160-E120 & 160 & 120 \\
      W220-E150 & 220 &150 \\
    \bottomrule
    \end{tabular}}
    \label{tab:complex_scenarios}
\end{table}

\section{Discussion} \label{Sec:Discussion}
The limited flexibility of current data-driven crowd simulation models across diverse scenarios represents a major constraint on their practical utility, underscoring the need for models with improved flexibility. Our results demonstrate that the proposed IVID model achieves good performance across four fundamental geometries (e.g., bottlenecks, corridors, corners, and T-junctions),  surpassing existing data-driven models in terms of scenario flexibility. \par

Bottlenecks constitute one of the most critical fundamental modules in pedestrian dynamics due to their impact on flow efficiency and safety risks. Although our original VID model \cite{Liang2024} performs well in corridors, corners and T-junctions, its inapplicability to bottlenecks represents a limitation. We hypothesize that this limitation arises from the failure of the original visual information extraction method, which is compromised by the frequent backward movement in bottlenecks resulting from high density. Consequently, we adjusted the visual information extraction method and incorporated exit information to enhance the detection of exits. The good performance of the IVID model in bottleneck scenarios validate our hypothesis and provide evidence of the core importance of visual information. \par

Developing a data-driven model with cross-scenario flexibility presents significant challenges, as it requires the accurate capture of the critical common information that guides pedestrians across various geometries. We argue that this core feature largely lies in visual information, which encompasses the scenario geometry and the pedestrian's position within the scene. The movement of pedestrians toward the scenario exit can be viewed as a dynamic process that involves repeatedly establishing and updating intermediate targets based on the currently available visual information and planning routes to reach these intermediate targets. The level of decision-making regarding these intermediate targets lies between the operational (e.g., determining movement at the next time step) and tactical (e.g., identifying the exit) levels \cite{HOOGENDOORN2004169, Haghani2016}. In essence, visual information serves as the key common factor enabling pedestrians to navigate effectively in diverse scenarios. Through the inherent perception of visual information, pedestrians can efficiently determine how to navigate turns at corners, identify their temporary target position, and conceive the route to reach that target. The success of our vision-based model across four geometries confirms its effectiveness in capturing these common factors, thereby validating the predominant importance of visual information, a feature often overlooked in prior data-driven models \cite{ZHAO2020, Mayi2016, ZHAO2021}. \par

Finally, we extended the application via a modular approach to simulate a composite scenario. While prior research has largely focused on simulating single modules \cite{ZHAO2021, Mayi2016, 10077452}, our work represents a step forward. In the absence of real trajectory data for the composite scenario, we evaluated performance using fundamental diagrams. The results show that the modular application of our IVID model yields fundamental diagrams highly consistent with controlled experiments, underscoring the substantial potential of data-driven approaches for modeling more complex real-world scenarios. \par

This study has several limitations. First, the model requires an initial 8-frame (lookback window) trajectory history, which may cause inconvenience for applications. This lookback window value follows previous research \cite{9043898, 7780479, DBLP}. Future studies could explore its sensitivity and investigate more efficient and user-friendly neural network architectures. Second, while knowledge-driven models derive theoretical generalizability from first principles and can generalize to previously unseen scenarios, our model's performance in unseen layouts remains an open question. This limitation is an inherent characteristic of the data-driven paradigm. Thus, our work advances multi-scenario flexibility within the data-driven paradigm, not a universal advantage. Third, we compare our model only with the classic SF model \cite{Helbing2000}. A broader comparison with contemporary models—necessitating careful calibration and scenario-specific adaptation—remains a direction for future work to more fully contextualize our model’s performance.

\par

\section{Conclusion} \label{Sec:Conclusion}
This paper presents a data-driven crowd simulation model with improved flexibility based on TCN. By effectively capturing visual and exit information, the model achieves good performance across four fundamental geometric modules: bottleneck, corridor, corner, and T-junction. We systematically evaluate the model against controlled experiments and the SF model \cite{Helbing2000} using multiple metrics, including trajectory analysis, fundamental diagrams, travel time, and distance errors. Results show that our model closely reproduces pedestrian trajectories and fundamental diagrams observed in experiments, while outperforming the SF model in both qualitative and quantitative comparisons across all four modules. Furthermore, we introduce a modular approach to simulate composite scenarios. The simulated trajectories and fundamental diagrams in the composite scenario align well with real-world observations, demonstrating the potential of data-driven methods for simulating complex pedestrian environments. The findings may inform future development of data-driven crowd simulation methods and promote broader application of data-driven approaches. \par

\par

\section*{Acknowledgments}
The work described in this paper was fully supported by a grant from CityU (Project No. DON\_RMG  9229030).

\bibliographystyle{elsarticle-num} 
\bibliography{ref}
\end{document}